\documentstyle[12pt]{article}

\begin{document}
\begin{titlepage}
\begin{flushleft}
Stockholm\\
USITP 95-2\\
February 1995\\
\end{flushleft}
\vspace{1cm}
\begin{center}
{\Large FORM GEOMETRY}\\
\ \\
{\Large AND THE 'tHOOFT-PLEBANSKI ACTION}\\
\vspace{2cm}
{\large Ingemar Bengtsson}\footnote{Email address: ingemar@vana.physto.se}\\
{\sl Fysikum\\
University of Stockholm\\
Box 6730, S-113 85 Stockholm, Sweden}\\
\vspace{3cm}
{\bf Abstract}\\
\end{center}
Riemannian geometry in four dimensions, including Einstein's equations, can
 be described by means of a connection that annihilates a triad of two-forms
 (rather than a tetrad of vector fields).
 Our treatment of the conformal factor of the metric differs from the
 original presentation of this result, due to 'tHooft. In the action
 the conformal factor now appears as a field to
be varied.

\end{titlepage}

\noindent {\bf 1. INTRODUCTION.}

\vspace*{5mm}

\noindent It is a fact that, in four dimensions, the set of all two-forms can
be divided into self-dual and anti-self dual ones. At first sight our fact
seems to be a boring algebraic one, but it is by no means so. Indeed this fact
somehow manages to serve as a corner stone of more than one imposing edifice -
Penrose's twistor theory, Ashtekar's variables for gravity, and Donaldson's
theory of four manifolds may serve as examples. Less grandly, but nevertheless
interestingly, 'tHooft \cite{'tHooft} pointed out that it allows a peculiar
variation of one of the fundamental theorems of Riemannian geometry, and it is
this observation which concerns us here.

The theorem (for $D$ dimensional spaces) is that one can use a $D$-ad of
 vector fields to relate the Riemann tensor and the curvature tensor of an
 SO($D$) connection that obeys

\begin{equation} D_{[{\alpha}}e_{{\beta}]I} = 0 \ . \end{equation}

\noindent More precisely, the Riemann tensor that one produces is the Riemann
tensor of the metric

\begin{equation} g_{{\alpha}{\beta}} = e_{{\alpha}I}e_{\beta}^{\ I} \
. \label{2} \end{equation}

\noindent 'tHooft's observation is that (for $D = 4$) we can we can use a
 triad of two-forms to  relate the Riemann tensor and the SO(3) curvature
 tensor of a connection that obeys

\begin{equation} \nabla_{[{\alpha}}{\Sigma}_{{\beta}{\gamma}]i} = 0 \
. \label{3} \end{equation}

\noindent What is now the analogue of eq. ({2})? The answer to this
 question is known, and will be reviewed in section 2.

In section 3 we make some observations about eq. (\ref{3}) which are of
interest in Yang-Mills theory; after all both the Bianchi identities and the
four dimensional Yang-Mills equations can be written in this form.

In section 4 we redo 'tHooft's analysis. Our treatment differs from his in the
way that we handle the conformal factor of the metric.

In section 5 we modify 'tHooft's formulation of the action
 principle for Einstein's equations, and bring it into line with the
 previous section.

We expect that the formalism discussed here can be useful in various
 contexts (indeed it can be seen as a natural outgrowth of
earlier alternatives to the Newman-Penrose formalism \cite{Israel}). We also
 harbour a suspicion that
the natural ``split'' of the metric into conformal factor and conformal
structure which happens here may turn out to be of considerable physical
interest. However, this is only a suspicion. Some comments on the formalism
are to be found in section 6.

\vspace*{1cm}

{\bf 2. THE SPACE OF TWO-FORMS.}

\vspace*{5mm}

\noindent Before we start, let me tell the reader that all my
${\epsilon}$-tensors take the values $\pm 1$ in every coordinate system,
hence they have non-zero density weights. To define them, all one needs is an
orientation of the space they live on. I never use a metric to raise or
lower indices on an ${\epsilon}$-tensor.

Now let {\bf V} be a four dimensional vector space and let {\bf W} be the
six dimensional vector space of two-forms on {\bf V}. We introduce  the
following
operations on {\bf W}:

\begin{eqnarray} \hspace*{1mm} \tilde{\Sigma}^{{\alpha}{\beta}} &=
& \frac{1}{2}{\epsilon}
^{{\alpha}{\beta}{\gamma}{\delta}}{\Sigma}_{{\gamma}{\delta}} \\
\ \nonumber \\
 \hspace*{1mm} *{\Sigma}_{{\alpha}{\beta}} &=
& \frac{1}{2\sqrt{\pm g}}g_{{\alpha}{\mu}}
g_{{\beta}{\nu}}{\epsilon}^{{\mu}{\nu}{\gamma}{\delta}}{\Sigma}_{{\gamma}
{\delta}} \ . \end{eqnarray}

\noindent The second definition requires a metric $g_{{\alpha}{\beta}}$ on
{\bf V}. (Actually a conformal structure is enough, since the definition
is conformally invariant.) The minus sign in $\sqrt{\pm g}$ is to be used
when this metric has Lorentzian signature. The eigenvectors of the star
operation are said to be self-dual or anti-self dual, depending on the
sign. Depending on the signature, a self-dual form obeys

\begin{eqnarray} (EK) \hspace*{15mm} *{\Sigma}_{{\alpha}{\beta}} &=
& {\Sigma}_{{\alpha}{\beta}} \\
\ \nonumber \\
(L) \hspace*{15mm} *{\Sigma}_{{\alpha}{\beta}} &=
& i{\Sigma}_{{\alpha}{\beta}} \ , \end{eqnarray}

\noindent where $(EK)$ in front of a formula means that is valid for Euclidean
 and Kleinian\footnote{Kleinian signature means that the number of plus and
minus signs are equal. The name is due to Gibbons.} signatures of
$g_{{\alpha}{\beta}}$, while $(L)$ stands for Lorentzian signature.

The space {\bf W} admits a natural metric, which is defined by

\begin{equation} <{\Sigma}, {\Sigma}> =
 {\Sigma}_{{\alpha}{\beta}}\tilde{\Sigma}^{{\alpha}{\beta}} \ . \end{equation}

\noindent We observe that this metric has Kleinian signature, and also that
the twiddle operation can be regarded as using this metric to raise indices
in {\bf W}.

The subspace of self-dual two-forms is three dimensional and will be denoted
${\bf W}^+$, while the subspace of anti-self dual forms is denoted
${\bf W}^-$. We assume that we have a basis ${\Sigma}_{{\alpha}{\beta}i}$
of ${\bf W}^+$ available, where the index $i$ runs from one to three. A
useful definition is

\begin{equation} m_{ij} = \ <{\Sigma}_i, {\Sigma}_j> \ . \end{equation}

\noindent This is a metric on ${\bf W}^+$. (But note carefully that it is a
scalar density of weight one under GL(4) transformations of {\bf V}.) Its
inverse will be denoted by $m^{ij}$;

\begin{equation} m_{ik}m^{kj} = {\delta}_i^j \ . \end{equation}

\noindent A particularly useful fact is that

\begin{equation} {\Sigma}_{{\alpha}{\gamma}i}
\tilde{\Sigma}^{{\gamma}{\beta}}_{\ \ j} + {\Sigma}_{{\alpha}{\gamma}j}
\tilde{\Sigma}^{{\gamma}{\beta}}_{\ \ i}  =
 - \frac{1}{2} m_{ij} {\delta}_{\alpha}^{\beta} \ . \end{equation}

So far this is very well known; first we introduce a metric on {\bf V}, then
the star operation is defined as a map from {\bf W} to {\bf W}, and finally
this map is used to define the self-dual subspace ${\bf W}^+$. What is less
well known is that the story can be told in reverse. We simply select any
three dimensional subspace of {\bf W}, define a star operator which turns
this subspace into the self-dual subspace of {\bf W}, and at the end use
this star operator to define a metric on {\bf V}. This metric will be
determined by the star operation uniquely up to conformal transformations;
the procedure provides an alternative way to characterize metrics on four
dimensional spaces. There are several points of view on this, and we refer
the reader to the literature for the details \cite{Urbantke} \cite{Donaldson}.

The central theorem in the subject is due to Urbantke:

\

\noindent \underline{Theorem} (Urbantke): The subspace ${\bf W}^+$ of
{\bf W} is the space of self-dual two-forms with respect to the metric

\begin{equation} g_{{\alpha}{\beta}} =
 \frac{8}{3}{\sigma}{\epsilon}^{ijk}{\Sigma}_{{\alpha}{\gamma}i}
\tilde{\Sigma}^{{\gamma}{\delta}}_{\ \ j}{\Sigma}_{{\delta}{\beta}k} \ ,
 \end{equation}

\noindent where the conformal factor ${\sigma}$ is arbitrary.

\noindent The signature of $g_{{\alpha}{\beta}}$ is

Euclidean if ${\Sigma}_{{\alpha}{\beta}i}$ are real and $m_{ij}$ has
 definite signature,

Kleinian if ${\Sigma}_{{\alpha}{\beta}i}$ are real and $m_{ij}$ has
 indefinite signature,

Lorentzian if ${\Sigma}_{{\alpha}{\beta}i}$ are complex and
$<{\Sigma}_i, \bar{\Sigma}_{j'}> = 0$.

\

\noindent With regard to the notation used here: Occasionally we use a
basis for the anti-self dual subspace ${\bf W}^-$, which is then denoted by
$\bar{\Sigma}_{i'}$. In the Lorentzian case these two-forms are
related by complex conjugation to the ${\Sigma}_i$'s, otherwise they are
unrelated objects.

The conformal factor ${\sigma}$ of Urbantke's metric is arbitrary (the
factor $8/3$ has been inserted for convenience), but it is needed to ensure
that the metric is invariant under GL(3) rotations in ${\bf W}^+$. Indeed
${\sigma}$ transforms as a scalar density of weight minus one under both
 GL(3) and GL(4). Finally it is
useful to know that

\begin{eqnarray} (EK) \hspace*{15mm} \sqrt{g} &=& 4{\sigma}^2m \\
\ \nonumber \\
(L) \hspace*{15mm} \sqrt{-g} &=& 4i{\sigma}^2m \ , \end{eqnarray}

\noindent where $m$ denotes the determinant of $m_{ij}$. From now on we
adopt Urbantke's metric in {\bf V}, so that $g_{{\alpha}{\beta}}$ is in
fact defined by our choice of the ${\Sigma}_i$'s, or exactly: The conformal
structure represented by $g_{{\alpha}{\beta}}$ is defined by our choice
of subspace ${\bf W}^+$.

A number of useful relations may now be derived, such as

\begin{eqnarray} (EK) \hspace*{13mm}
{\Sigma}_{{\alpha}{\gamma}i}\tilde{\Sigma}_{\ {\beta}j}^{\gamma} &=
& - \frac{1}{4}m_{ij}g_{{\alpha}{\beta}} -
{\sigma}m{\epsilon}_{ijk}{\Sigma}^{\ \ k}_{{\alpha}{\beta}} \\
\ \nonumber \\
(L) \hspace*{13mm} {\Sigma}_{{\alpha}{\gamma}i}
\tilde{\Sigma}_{\ {\beta}j}^{\gamma} &=
& - \frac{1}{4}m_{ij}g_{{\alpha}{\beta}} +
{\sigma}m{\epsilon}_{ijk}{\Sigma}^{\ \ k}_{{\alpha}{\beta}} \label{16} \\
\ \nonumber \\
{\Sigma}_{[{\alpha}|{\gamma}i}\bar{\Sigma}^{\gamma}_{\ {\beta}]j'} &=
& 0 \\
\ \nonumber \\
(EK) \hspace*{1cm}
 {\Sigma}_{{\alpha}{\beta}i}\tilde{\Sigma}^{{\gamma}{\delta}i}
 &=& \frac{1}{2}(1 + *)_{{\alpha}{\beta}}^{\ \ \ {\gamma}{\delta}} \\
\ \nonumber \\
(L) \hspace*{1cm}
 {\Sigma}_{{\alpha}{\beta}i}\tilde{\Sigma}^{{\gamma}{\delta}i}
 &=& \frac{1}{2}(1 - i*)_{{\alpha}{\beta}}^{\ \ \ {\gamma}{\delta}},
\end{eqnarray}

\noindent where we have used $m_{ij}$ and $g_{{\alpha}{\beta}}$ to raise and
 lower Latin and Greek indices on the two-forms, respectively. (We will do
 this without comment in the sequel.)

Under certain conditions, the formalism that we are developing reduces to
 the familiar tetrad formalism based on vectors in {\bf V}. A precise
 statement \cite{Capovilla} is the following:

\

\noindent \underline{Theorem} (Capovilla et al.): The condition

\begin{equation} m_{ij} \propto {\delta}_{ij} \end{equation}

\noindent guarantees the existence of a tetrad of vectors $e_{{\alpha}I}$
 such that

\begin{equation} {\Sigma}_{{\alpha}{\beta}i} = e_{{\alpha}{\beta}i}
\hspace*{2cm}
g_{{\alpha}{\beta}} = e_{{\alpha}I}e_{\beta}^{\ I} \ , \end{equation}

\noindent where $e_{{\alpha}{\beta}i}$ denotes the self-dual part of the
 two-form

\begin{equation} e_{{\alpha}{\beta}}^{\ \ IJ} \equiv
 e_{[{\alpha}}^{\ \ I}e_{{\beta}]}^{\ \ J} \ . \end{equation}

\

As an application \cite{Israel} of the form formalism we consider the
 Riemann tensor (or any tensor with the same index symmetries). It is
 elementary to show that the Riemann tensor can be expressed as

\begin{equation} R_{{\alpha}{\beta}{\gamma}{\delta}} =
 {\Sigma}_{{\alpha}{\beta}i}r^{ij}{\Sigma}_{{\gamma}{\delta}j} +
 \bar{\Sigma}_{{\alpha}{\beta}i'}\bar{r}^{i'j'}
\bar{\Sigma}_{{\gamma}{\delta}j'} + {\Sigma}_{{\alpha}{\beta}i}s^{ij'}
\bar{\Sigma}_{{\gamma}{\delta}j'} +
\bar{\Sigma}_{{\alpha}{\beta}i'}\bar{s}^{i'j}{\Sigma}_{{\gamma}{\delta}j}
 \ , \label{23} \end{equation}

\noindent where

\begin{equation} r^{ij} = r^{ji}
\hspace*{1cm} \bar{r}^{i'j'} = \bar{r}^{j'i'}
\hspace*{1cm} \bar{s}^{j'i} = s^{ij'} \end{equation}

\noindent and

\begin{equation} m_{ij}r^{ij} + \bar{m}_{i'j'}\bar{r}^{i'j'} = 0 \ .
\end{equation}

\noindent The last condition comes from the cyclic property of the Riemann
 tensor. In the Lorentzian case, the bar denotes complex conjugation. For
 the traceless Ricci tensor and the curvature scalar we find

\begin{eqnarray} R_{{\alpha}{\beta}} - \frac{1}{4}Rg_{{\alpha}{\beta}} &=
& - s^{ij'}({\Sigma}_{{\alpha}{\gamma}i}\bar{\Sigma}^{\gamma}_{\ {\beta}j'}
+ {\Sigma}_{{\beta}{\gamma}i}\bar{\Sigma}^{\gamma}_{\ {\alpha}j'}) \\
\ \nonumber \\
(EK) \hspace*{2cm} \sqrt{g}R &=& 2m_{ij}r^{ij} \\
\ \nonumber \\
(L) \hspace*{2cm} \sqrt{-g}R &=& - 2im_{ij}r^{ij} \ . \end{eqnarray}

\noindent These formul\ae \ will be used below.

\vspace*{1cm}

{\bf 3. YANG-MILLS GEOMETRY.}

\vspace*{5mm}

\noindent Our first exercise is to solve the equation

\begin{equation} D_{[{\alpha}}{\Sigma}_{{\beta}{\gamma}]i} =
0 \ , \label{2.1} \end{equation}

\noindent where

\begin{equation} D_{\alpha}{\Sigma}_{{\beta}{\gamma}i} =
\partial_{\alpha}{\Sigma}_{{\beta}{\gamma}} +
 {\epsilon}_{ijk}A_{{\alpha}j}{\Sigma}_{{\beta}{\gamma}k} \ , \end{equation}

\noindent and the GL(3) invariance is broken down to SO(3) since we have
 chosen the Kronecker delta to raise and lower the internal indices of the
 two-forms. This is twelve equations for twelve unknowns.

The exercise is quite straightforward. For definiteness we choose the
conventions that lead to a Euclidean signature for the metric. We find
the result

\begin{equation} A_{{\alpha}i} =
 - 2t_{{\alpha}{\beta}{\gamma}ij}\tilde{\Sigma}^{{\beta}{\gamma}}_{\ \ k}
m^{jk} - 2{\Sigma}_{{\alpha}{\beta}i}
\tilde{\Sigma}_{\ \ j}^{{\beta}{\gamma}}t_{{\gamma}{\delta}{\sigma}km}
\tilde{\Sigma}^{{\delta}{\sigma}}_{\ \ n}m^{mn}m^{jk} \ , \end{equation}

\noindent where

\begin{equation} t_{{\alpha}{\beta}{\gamma}ij} =
{\epsilon}_{ijk}(\partial_{\alpha}{\Sigma}_{{\beta}{\gamma}k} +
 \partial_{\gamma}{\Sigma}_{{\alpha}{\beta}k} +
 \partial_{\beta}{\Sigma}_{{\gamma}{\alpha}k}) \ . \end{equation}

The result is of some interest for Yang-Mills theory, in two ways. First
 we choose

\begin{equation} {\Sigma}_{{\alpha}{\beta}i} =
F_{{\alpha}{\beta}i} \ . \end{equation}

\noindent Then what we have done is that we have solved the Bianchi
 identities for the connection in terms of the field strength, under the
 assumption that the latter is non-degenerate in the sense that
 $\det m_{ij} \neq 0$. Hence the Wu-Yang ambiguity is quite ``mild'' in
 four dimensions.

Alternatively we can choose

\begin{equation} {\Sigma}_{{\alpha}{\beta}i} = \star F_{{\alpha}{\beta}i}
 \ , \end{equation}

\noindent where the $\star $ denotes the star operator defined using the
 physical space-time metric, which is not the metric that we can construct
 from the two-forms. Then what we have done is to solve the Yang-Mills
 equations for the connection as a function of the field strength and the
 physical metric, which we denote as $h_{{\alpha}{\beta}}$, again under a
 non-degeneracy condition on the field strength. Where did the dynamics go?
 Actually it is still there, in the equation

\begin{equation} F_{{\alpha}{\beta}i} = F_{{\alpha}{\beta}i}(A(F,h)) \ .
 \label{3.6} \end{equation}

\noindent This equation is not a pleasant one to analyze, but if we are
 prepared to disregard the Wu-Yang ambiguity it may at least be regarded
 as an interesting curiousity, and perhaps more.

In three dimensions one can not use the Bianchi identities in the same way
 - in fact there is then only one Bianchi identity, and the Wu-Yang
 ambiguity becomes more serious. If this is disregarded one can solve the
 Yang-Mills equations in a manner which is analogous to the above. Moreover,
 in three dimensions the analogue of eq. (\ref{3.6}) can be written as an
 equation for the Ricci tensor of the ``Yang-Mills metric'' that one can
 form from the field strength \cite{Lunev}. I do not know whether a similar
 interpretation can be made for eq. (\ref{3.6}).

\vspace*{1cm}

{\bf 4. RIEMANNIAN GEOMETRY.}

\vspace*{5mm}

\noindent It is crucial to have a direct relation between the curvature
 tensor of a connection acting on self-dual two-forms, on the one hand,
 and the Riemann tensor of the Urbantke metric on the other. Provided that
 we make some slight changes in the definitions of the previous section,
 such a relation can be indeed be found. Specifically, we will allow the
 connection to take values in the Lie algebra of GL(3), and not just in
 any preassigned SO(3) subspace.

Since the metric will eventually be identified with the physical metric, we
choose the conventions appropriate to Lorentzian space-times.  We also adopt
 the convention that GL(3) indices are raised and lowered with the metric
$m_{ij}$. And so we impose the condition

\begin{equation}
D_{[{\alpha}}{\Sigma}_{{\beta}{\gamma}]i} =
 \partial_{[{\alpha}}{\Sigma}_{{\beta}{\gamma}]i} +
{\cal A}_{[{\alpha}|i}^{\ \ \ j}{\Sigma}_{{\beta}{\gamma}]j} = 0 \ .
\label{36} \end{equation}

These are only twelve equations, so it is hard to see how we can solve them
 for a  GL(3) valued connection.  Let us therefore postpone this question
 and go to the next step, which is to introduce an affine connection
 through the equation

\begin{equation}
{\Gamma}_{{\alpha}{\beta}}^{\ \ {\delta}}{\Sigma}_{{\delta}{\gamma}i} -
{\Gamma}_{{\alpha}{\gamma}}^{\ \ {\delta}}{\Sigma}_{{\delta}{\beta}i} =
D_{\alpha}{\Sigma}_{{\beta}{\gamma}i} \ .
\label{37} \end{equation}

\noindent The condition (\ref{36}) implies that the affine connection is
 symmetric. If we then count components, we have 40 unknowns. Since a
 projection to the self-dual subspace is involved - this is not quite self
 evident, but it follows because the projection operator is a GL(3)
 scalar - the number of equations is only 36, so that eq. ({\ref{37})
 underdetermines the affine
connection.

At this point we recall the arbitrary factor in Urbantke's metric. We can
 raise the number of equations for the affine connection to 40 by
 introducing a field ${\sigma}(x)$ which transforms as a scalar density of
 weight minus one under both GL(3) and GL(4). Our final claim is that,
 once eq. (\ref{36}) is imposed, an affine connection is defined by the
 equations

\begin{eqnarray} \nabla_{\alpha}{\Sigma}_{{\beta}{\gamma}i} &=& 0
\label{38} \\
\ \nonumber \\
\nabla_{\alpha}{\sigma} &=& 0 \ ,
\label{39} \end{eqnarray}

\noindent where

\begin{eqnarray} \nabla_{\alpha}{\Sigma}_{{\beta}{\gamma}i} &=&
{\cal D}_{\alpha}{\Sigma}_{{\beta}{\gamma}i} +
{\cal A}_{{\alpha}i}^{\ \ j}{\Sigma}_{{\beta}{\gamma}j} =
\partial_{\alpha}{\Sigma}_{{\beta}{\gamma}i} -
{\Gamma}_{{\alpha}{\beta}}^{\ \ {\delta}}{\Sigma}_{{\delta}{\gamma}i} -
{\Gamma}_{{\alpha}{\gamma}}^{\ \ {\delta}}{\Sigma}_{{\beta}{\delta}i} +
{\cal A}_{{\alpha}i}^{\ \ j}{\Sigma}_{{\beta}{\gamma}j} \ , \\
\ \nonumber \\
\nabla_{\alpha}{\sigma} &=& \partial_{\alpha}{\sigma} +
 {\Gamma}_{{\alpha}{\gamma}}^{\ \ {\gamma}}{\sigma} -
 {\cal A}_{{\alpha}i}^{\ \ i}{\sigma} \ . \end{eqnarray}

\noindent Note that there are by now three covariant derivatives
in the game, $D_{\alpha}$, ${\cal D}_{\alpha}$ and ${\nabla}_{\alpha}$.

These are the form compatibility conditions, and we must now verify that we
can solve them for the affine connection. To do this we introduce Urbantke's
 expression for the metric $g_{{\alpha}{\beta}}$, and use the fact that it is
a GL(3) scalar to deduce that

\begin{equation} {\cal D}_{\gamma}g_{{\alpha}{\beta}} =
 \nabla_{\gamma}g_{{\alpha}{\beta}}({\sigma},{\Sigma}) = 0 \ . \end{equation}

\noindent Hence the affine connection is metric compatible, and can be
 expressed as Christoffel symbols in the usual way.

With eq. (\ref{38}) and the solution for the affine connection in hand, it
 is of course straightforward to solve for the GL(3) valued connection as
 a function of the two-forms and ${\sigma}$. We obtain

\begin{equation} {\cal A}_{{\alpha}i}^{\ \ j} =
- \tilde{\Sigma}^{{\beta}{\gamma}j}
{\cal D}_{\alpha}{\Sigma}_{{\beta}{\gamma}i} \ .
\end{equation}

\noindent So this problem is solved.

Our next goal, and our main goal, is to relate the curvature tensors. From

\begin{equation} 0 =
 [\nabla_{\alpha}, \nabla_{\beta}]{\Sigma}_{{\gamma}{\delta}i} =
 R_{{\alpha}{\beta}{\gamma}}^{\ \ \ \ {\sigma}}{\Sigma}_{{\sigma}{\delta}i} -
 R_{{\alpha}{\beta}{\delta}}^{\ \ \ \ {\sigma}}{\Sigma}_{{\sigma}{\gamma}i} +
 {\cal F}_{{\alpha}{\beta}i}^{\ \ \ j}{\Sigma}_{{\gamma}{\delta}j}
 \end{equation}

\noindent we may deduce that

\begin{equation} {\cal F}_{{\alpha}{\beta}ij} =
 - 2{\sigma}m{\epsilon}_{ijk}R_{{\alpha}{\beta}}^{\ \ \ {\gamma}{\delta}}
{\Sigma}_{{\gamma}{\delta}}^{\ \ k} =
\frac{1}{2{\sigma}}{\epsilon}_{ijk}(r^{km}{\Sigma}_{{\alpha}{\beta}m} +
s^{km'}\bar{\Sigma}_{{\alpha}{\beta}m'}) \ .
\label{47} \end{equation}

\noindent (In the second step we made use of the notation for the
 Riemann tensor that was introduced in section 2.) So we see that the
 GL(3) curvature lies in an SO(3) subalgebra, whatever
the choice of ${\Sigma}_i$'s, and moreover that it is simply related to
the self-dual part of the Riemann tensor. Conversely, we may express the
latter in terms of the former, namely through the equations

\begin{eqnarray} r^{ij} &=
& {\sigma}{\epsilon}^{imn}\tilde{\Sigma}^{{\alpha}{\beta}j}
{\cal F}_{{\alpha}{\beta}mn} \\
\ \nonumber \\
s^{ij'} &=
& {\sigma}{\epsilon}^{imn}\tilde{\bar{\Sigma}}^{{\alpha}{\beta}j'}
{\cal F}_{{\alpha}{\beta}mn} \\
\ \nonumber \\
\sqrt{-g}R &=
& - 2i{\sigma}{\epsilon}^{ijk}\tilde{\Sigma}^{{\alpha}{\beta}}_{\ \ i}
{\cal F}_{{\alpha}{\beta}jk} \ . \end{eqnarray}

With this, our proof of 'tHooft's form version of the fundamental theorem
 of Riemannian geometry is complete. A few remarks suggest themselves;
 first of all the part played by the conformal factor ${\sigma}$ in the
 proof is worth watching. Second, in the end it is not surprising that we
 were able to solve the twelve equations (\ref{36}) for the connection,
 because it turned out to be an SO(3) connection after all; specifically
 we see that

\begin{equation} D_{\alpha}({\sigma}m_{ij}) =
 \nabla_{\alpha}({\sigma}m_{ij}) =0 \ . \end{equation}

\noindent This defines the metric which selects the relevant SO(3) subspace
 of GL(3). Third, if we refer back to eq. (\ref{23}) we see that
 we have not been able to express the entire Riemann tensor as a function
 of the curvature tensor ${\cal F}_{{\alpha}{\beta}}$; we are missing
 the traceless part of the matrix $\bar{r}^{i'j'}$, which is the same
 thing as the anti-self dual part of the Weyl tensor. This happened
 because a self-dual projection was built into eq. (\ref{37}); having
 solved for the affine connection we can of course use the resulting
 expression to act on anti-self dual forms as well, but this does involve
 a choice. In the real Lorentzian case the anti-self dual Weyl tensor
 can be reached from the self-dual Weyl tensor through complex conjugation,
 but otherwise they are algebraically independent objects. (And the
 Lorentzian reality conditions are awkward to impose.)

\vspace*{1cm}

{\bf 5. GENERAL RELATIVITY.}

\vspace*{5mm}

\noindent 'tHooft's paper goes on to show how Einstein's equations can
 be derived from an action that is a functional of a  connection and a
 triad of two-forms. We will repeat his construction here, with the
 minor changes caused by the deviations from his treatment that we have
 already made.

It should occasion no surprise that the action is

\begin{equation} S[{\cal A}, {\Sigma}, {\sigma}] = \int \sqrt{-g}
(R - 2{\lambda}) =
-2i \int ({\sigma}{\epsilon}^{ijk}\tilde{\Sigma}^{{\alpha}{\beta}}_{\ \ i}
{\cal F}_{{\alpha}{\beta}jk} + 4{\lambda}{\sigma}^2m) \ .
\end{equation}

\noindent We observe that the trace of the connection drops out of the
 action, so that this is a functional of an SL(3) valued connection only,
 quite in accordance with ref. \cite{'tHooft}.

Varying the action with respect to the connection, we find (after minor
 manipulations) the equation

\begin{equation} {\epsilon}^{ijk}{\sigma}\hat{D}_{\beta}
\tilde{\Sigma}^{{\alpha}{\beta}}_{\ \ k} =
 - {\epsilon}^{imn}\tilde{\Sigma}^{{\alpha}{\beta}}_{\ \ n}m^{jk}
\hat{D}_{\beta}({\sigma}m_{km}) \ , \label{49} \end{equation}

\noindent where $\hat{D}_{\alpha}$ is an SL(3) covariant derivative. It
 will require some effort to extract the content of this equation; we begin
 with the observation that the equation remains true if the SL(3) covariant
 derivative is replaced by a GL(3) covariant derivative $D_{\alpha}$. The
 trace of the connection is then at our disposal, and we are free to define
 it through the equation

\begin{equation} m^{ij}D_{\alpha}({\sigma}m_{ij}) = 0 \ . \end{equation}

\noindent From now on we assume that this has been done. Then the next step
 is to write the symmetric part of eq. (\ref{49}) in the form

\begin{equation} M_{{\alpha}ij}^{\ \ \ {\beta}kl}D_{\beta}({\sigma}m_{kl}) =
 0 \ , \end{equation}

\noindent where we regard $M$ as a matrix acting on vectors that are
 symmetric and traceless in their Latin indices. We need to show that this
 matrix is invertible, but we do not necessarily have to invert it. Now it
 is not difficult - using eq. (\ref{16}) - to show that

\begin{equation} M^2 + \frac{1}{2{\sigma}}M =
 \frac{3i}{2{\sigma}^2}{\bf 1} \ . \end{equation}

\noindent By going to Jordan's canonical form, we see that this matrix
 equation does not allow $M$ to have any zero eigenvalues, therefore
 it is indeed invertible, and we may conclude that

\begin{equation} D_{\alpha}({\sigma}m_{ij}) = 0 \ . \label{50} \end{equation}

\noindent Then the anti-symmetric part of eq. (\ref{49}) gives

\begin{equation} D_{\beta}\tilde{\Sigma}^{{\alpha}{\beta}}_{\ \ \ i} = 0 \ .
\end{equation}

\noindent This is precisely eq. (\ref{36}), and when taken together these
 equations are equivalent to the form compatibility conditions (\ref{38}
 - \ref{39}). Therefore the content of these field equations is given by
 the formula (\ref{47}), the one that relates the curvature tensor to the
 self-dual part of the Riemann tensor of Urbantke's metric.

When we vary the action with respect to $\tilde{\Sigma}^{{\alpha}{\beta}}
_{\ \ \ i}$, and use eq. (\ref{47}) for the curvature tensor in the
 resulting equation, we obtain

\begin{eqnarray}
& (r_j^{\ j} + 8{\lambda}{\sigma}^2m){\Sigma}_{{\alpha}{\beta}}^{\ \ i} +
s^{ij'}\bar{\Sigma}_{{\alpha}{\beta}j'} = 0 \\
& \hspace*{8cm} \Leftrightarrow  \nonumber \\
& R_{{\alpha}{\beta}} = {\lambda}g_{{\alpha}{\beta}} \ . \end{eqnarray}

\noindent This is Einstein's equations for an arbitrary cosmological
 constant ${\lambda}$.

Finally, variation with respect to ${\sigma}$ gives nothing new. This
 concludes the demonstration that 'tHooft's action gives Einstein's
 equations.

Coupling to electromagnetism is straightforward; the only trick employed is
 the addition of a surface term to the matter action, which is

\begin{eqnarray} S[{\cal A}, {\Sigma}, {\sigma}, A] &=
& \int \sqrt{-g}(R - 2{\lambda} - \frac{1}{4}F^{{\alpha}{\beta}}(1 - i*)
F_{{\alpha}{\beta}}) = \nonumber \\
\ \\
&=& -2i \int ({\sigma}{\epsilon}^{ijk}\tilde{\Sigma}^{{\alpha}{\beta}}_{\ \ i}
{\cal F}_{{\alpha}{\beta}jk} + 4{\lambda}{\sigma}^2m -
 \frac{1}{4}F_{{\alpha}{\beta}}\tilde{\Sigma}^{{\alpha}{\beta}}_{\ \ i}
\tilde{\Sigma}^{{\gamma}{\delta}i}F_{{\gamma}{\delta}} ) \ .
\nonumber \end{eqnarray}

\noindent Variation with respect to the vector potential gives Maxwell's
 equations, and variation with respect to the two-form triad gives the
 remaining Einstein-Maxwell equations in the form

\begin{equation}
 (r_j^{\ j} + 8{\lambda}{\sigma}^2m){\Sigma}_{{\alpha}{\beta}}^{\ \ i} +
(s^{ij'} - \frac{1}{2}F^i\bar{F}^{j'})\bar{\Sigma}_{{\alpha}{\beta}j'} = 0 \
 , \end{equation}

\noindent where we made the obvious definition

\begin{equation} F_{{\alpha}{\beta}} = F^i{\Sigma}_{{\alpha}{\beta}i} +
 \bar{F}^{i'}\bar{\Sigma}_{{\alpha}{\beta}i'} \ . \end{equation}

\noindent Coupling to spinorial matter requires more elaborate measures,
 since SL(3) does not have spinorial representations.

As a matter of fact 'tHooft's action is closely related to an action first
studied in the seventies by Plebanski \cite{Plebanski}. Plebanski's action
differs from 'tHooft's only in that Plebanski adds the constraint

\begin{equation} {\sigma}m_{ij} = {\delta}_{ij} \end{equation}

\noindent to the action by means of a Lagrange multiplier. Because of the
 theorem by Capovilla et al. that we quoted earlier, the formalism then
 rapidly collapses to the familiar tetrad formalism. Moreover it is known
 that Ashtekar's Hamiltonian formulation of Einstein's equations can be
 obtained from the Plebanski action in a few easy steps \cite{Capovilla}.
 With Plebanski's constraint added, we are obviously dealing with an SO(3)
 connection only. This is actually the case also in the more general
 setting considered by 'tHooft, with the interesting difference that the
 relevant SO(3) subspace of GL(3) is then determined dynamically by the
 field equations, rather than imposed from the outside.

\vspace*{1cm}

{\bf 6. DISCUSSION.}

\vspace*{5mm}

\noindent Having introduced four different covariant derivatives I may have
 lost the reader, so let me summarize the results before discussing them. We
 start with an action that depends on an SL(3) connection:

\begin{equation} S = -2i \int {\sigma}{\epsilon}^{ijm}m_{mk}
\tilde{\Sigma}^{{\alpha}{\beta}}_{\ \ i}
{\cal F}_{{\alpha}{\beta}j}^{\ \ \ \ k} \ .
\end{equation}

\noindent Varying the action with respect to the connection, one finds that
 the field equations can be rewritten using a GL(3) covariant derivative
 $D_{\alpha}$, with the trace of the connection chosen in a particular
 way, so that the equations take the form

\begin{equation} D_{\beta}\tilde{\Sigma}^{{\alpha}{\beta}}_{\ \ \ i} = 0
\hspace*{3cm} D_{\alpha}({\sigma}m_{ij}) = 0 \ . \end{equation}

\noindent These conditions guarantee that the covariant derivative
 can be further extended to a GL(4) covariant derivative $\nabla_{\alpha}$,
 defined using a symmetric affine connection, such that

\begin{equation} \nabla_{\alpha}{\Sigma}_{{\beta}{\gamma}i} = 0
\hspace*{2cm} \nabla_{\alpha}{\sigma} = 0 \ . \hspace*{1cm} \label{64}
 \end{equation}

\noindent In their turn these conditions guarantee that the derivative
 ${\nabla}_{\alpha}$ is compatible with the metric

\begin{equation} g_{{\alpha}{\beta}} =
 \frac{8}{3}{\sigma}{\epsilon}^{ijk}{\Sigma}_{{\alpha}{\gamma}i}
\tilde{\Sigma}^{{\gamma}{\delta}}_{\ \ j}{\Sigma}_{{\delta}{\beta}k}
 \hspace*{5mm} \label{65} \end{equation}

\noindent - which is, up to a factor, the unique  metric tensor with
 respect to which the ${\Sigma}_i$'s are self-dual. Now eq. ({\ref{64})
 allows us to express the curvature tensor of the original connection in
 terms of the self-dual Riemann tensor of the metric (\ref{65}). If we go
 back to the action, vary with respect to the two-form, and insert the
 expression for the curvature tensor that we just obtained, we find
 Einstein's equations. Variation with respect to ${\sigma}$ gives nothing
 new. A cosmological constant and couplings to matter can be added without
 any ado.

An obvious drawback which our formalism shares with all ``chiral''
 formalisms for gravity (such as Ashtekar's variables) is that the
 Lorentzian reality conditions are awkward to impose. There are also
 some obvious strengths; formalisms that are related to ours have been
 extensively used in many problems such as classifying geometries and
 the like (see ref. \cite{Israel} for a review). However, the feature
 that we wish to stress is the clean separation of the metric into
 conformal structure and conformal factor which is achieved here, through
 a peculiarly four dimensional mechanism.

The way this happens is somewhat analogous to the appearance of the
 ``complexion'' in Rainich's formulation of electrodynamics \cite{Rainich};
 there is a factor left undetermined in the algebraic part of the
 discussion, which then turns into a field in the differential part. In
 our case it is essential that the field ${\sigma}$ carries non-zero
 density weight with respect to both GL(3) and GL(4). A certain scalar
 density plays a similar role also in the CDJ action for gravity
\cite{CDJ}, which is a functional of this field and the self-dual spin
 connection alone. However, the CDJ action suffers from two drawbacks which
 are not shared by our action; it breaks down for certain algebraically
 special field configurations, and it is hopelessly complicated for almost
 anything except pure gravity with vanishing cosmological constant
 \cite{Peldan}.

A feature peculiar to the present formalism is that the variation of the
 action with respect to the field ${\sigma}$ - to which we are loosely
 referring as ``the conformal factor'' - does not add further content to
 the field equations. It might therefore seem to be a very innocent
 bystander in the theory. Nevertheless I suspect that the present
 formalism can be used to illuminate a wide range of problems in relativity
 where conformal transformations are being made.

\

\

\noindent {\bf Acknowledgements:}

\

\noindent I thank Peter Peld\'{a}n for a critical reading of the manuscript,
and the NFR for financial support.

\end{document}